\documentclass[superscriptaddress]{revtex4}
\usepackage{amsmath}
\usepackage{graphicx}
\usepackage{floatflt,epsfig}
\begin{document}

\def\qgp{quark--gluon--plasma}
\def\njl{Nambu--Jona--Lasinio}
\def\cs{chiral symmetry}
\def\ii{\'{\i}}
\def\d{\mbox{d}}

\title{Warm stellar matter with neutrino trapping}

\author{D.P.Menezes}
\affiliation{Depto de F\'{\i}sica - CFM - Universidade Federal de Santa
Catarina  Florian\'opolis - SC - CP. 476 - CEP 88.040 - 900 - Brazil}

\author{C. Provid\^encia}
\affiliation{Centro de F\ii sica Te\'orica - Dep. de F\ii sica - 
Universidade de Coimbra - P-3004 - 516 Coimbra - Portugal}

\begin{abstract}

The properties of hybrid stars formed by hadronic and quark matter
in $\beta$-equilibrium at fixed entropies are described by appropriate 
equations of state (EOS) in the framework of relativistic mean-field
theory. In this work we include the possibility of neutrino trapped EOS
and compare the star properties with the ones obtained after deleptonization,
when neutrinos have already diffused out. We use the nonlinear Walecka model 
for the hadron matter with two different sets for the hyperon couplings and the MIT Bag and the Nambu-Jona-Lasinio models for 
the quark matter. The phase transition to a deconfined quark phase 
is investigated. Depending on the model and the parameter set used, 
the mixed phase may or may not exist in the EOS at high densities. 

The star properties are calculated for each equation of state.  
The maximum mass stellar configurations obtained within the NJL have larger 
masses than the ones obtained within the Bag model. 
The Bag  model predicts a mixed phase in the interior of the 
most massive stable stars  while, depending on the hyperon couplings, the NJL 
model predicts a mixed phase or pure quark matter. Comparing with neutrino 
free stars, the maximum allowed baryonic masses for
protoneutron stars are $\sim 0.4 M_\odot$ larger for the Bag model
and  $\sim 0.1 M_\odot$ larger for the NJL model when neutrino trapping is 
imposed.
\end{abstract}

\maketitle

\vspace{0.50cm}
PACS number(s): {95.30.Tg, 21.65.+f, 21.30.-x}
\vspace{0.50cm}

\section{Introduction}

Protoneutron stars appear as an outcome of the gravitational colapse of a 
massive star. During its early evolution a protoneutron star with an entropy
per baryon of the order of 1 to 2 contains trapped neutrinos. After 10 to 20 
seconds, the star stabilizes at practically zero temperature and no trapped 
neutrinos are left \cite{prak97}. The structure of compact stars is 
characterized by its mass and radius, which are obtained from appropriate 
equations of state (EOS) at densities about one order of magnitude higher than 
those observed in ordinary nuclei. EOS can be derived either from relativistic 
or potential models. The last ones are normally developed within a non 
relativistic formalism \cite{nonrel} and most of them are only suited at low densities  
because the EOS becomes acausal, i.e., the speed of sound exceeds the speed 
of light at densities which are relevant for neutron and protoneutron stars. 
Moreover, these models may lead to symmetry energies that decrease much more 
than expected beyond three times saturation density and this is a serious 
deficiency for neutron stars, which are highly asymmetric systems. 
In  relativistic models these problems are not present.
In this work we
investigate the properties of hybrid stars formed by hadronic matter at low 
densities, mixed matter at intermediate densities and quark matter at high 
densities. We use the nonlinear Walecka model for the hadron phase 
\cite{walecka,bb} and the MIT Bag \cite{bag} and the Nambu-Jona-Lasinio 
\cite{NJL,Klevansky92} models for the quark matter. 

In a previous work \cite{recente} we have used the same formalism in order to 
study the properties of hybrid stars obtained from EOS at different 
temperatures. We have checked that the maximum masses of hybrid stars obtained
with the Bag model are of the order of $\sim 1.6$ M$_{\bigodot}$,  smaller 
than the maximum masses obtained within the NJL model, 
$\sim 1.9$M$_{\bigodot}$.  
Even with a transition to a deconfinement phase the masses predicted can be quite high. However, the effect of temperature in the maximum masses allowed is 
not strong.  The central energy densities, though, decrease with temperature.
Some of the features exhibited for the Bag model were quite different from the
ones obtained with the NJL model, due to the chiral conservation mechanism 
implicit in the latter one.

In this work, we verify the importance of including trapped neutrinos and 
consider entropies from zero to 2 Boltzmann units. Most of the formulae used 
for the lagrangian densities, energies, pressures, partition functions, etc,
are not shown in this paper because they are standard equations and can be 
seen, among others, in Ref. \cite{recente}. We compare the properties of warm 
stars obtained within the NJL model and the MIT bag model, namely, 
strangeness content, neutrino fraction,  onset of hyperons, mixed phase and 
quark phase, maximum allowed  mass and interior composition.

The present paper is organized as follows: in Sec. II the lagrangian 
densities of the models used in the hadronic and quark matter are described, 
important relations and parameter sets are displayed and the mixed phase is 
implemented. The results are shown and discussed in Sec. III and in 
the last Sec. some remarks are made. 

\section{Formalism}

For the hadron phase we have used the nonlinear Walecka model with the 
inclusion of hyperons. The lagrangian density of the model reads:

$$
{\cal L}_{NLWM}= \sum_B \bar \psi_B \left[\gamma_\mu\left (i\partial^{\mu}
-g_{vB} V^{\mu}- g_{\rho B} {\mathbf t} \cdot {\mathbf b}^\mu \right) 
-(M_B-g_{s B} \phi)\right]\psi_B$$ 
$$+ \frac{1}{2}(\partial_{\mu}\phi\partial^{\mu}\phi
-m_s^2 \phi^2) - \frac{1}{3!}\kappa \phi^3 -\frac{1}{4!}\lambda
\phi^4$$
$$-\frac{1}{4}\Omega_{\mu\nu}\Omega^{\mu\nu}+\frac{1}{2}
m_v^2 V_{\mu}V^{\mu} + \frac{1}{4!}\xi g_v^4 (V_{\mu}V^{\mu})^2
-\frac{1}{4}{\mathbf B}_{\mu\nu}\cdot{\mathbf B}^{\mu\nu}+\frac{1}{2}
m_\rho^2 {\mathbf b}_{\mu}\cdot {\mathbf b}^{\mu}
$$
\begin{equation}
+\sum_l \bar \psi_l \left(i \gamma_\mu \partial^{\mu}-
m_l\right)\psi_l,
\end{equation}
with $\sum_B$ extending over the eight baryons,
$g_{s B}=x_{s B}~ g_s,~~g_{v B}=x_{v B}~ g_v,~~g_{\rho B}=x_{\rho B}~ 
g_{\rho}$
and $x_{s B}$, $x_{v B}$ and $x_{\rho B}$ are equal to $1$ for the nucleons and
acquire different values in different parametrizations for the other baryons,
$\Omega_{\mu\nu}=\partial_{\mu}V_{\nu}-\partial_{\nu}V_{\mu}$,
${\mathbf B}_{\mu\nu}=\partial_{\mu}{\mathbf b}_{\nu}-\partial_{\nu} 
{\mathbf b}_{\mu}
- g_\rho ({\mathbf b}_\mu \times {\mathbf b}_\nu)$
and ${\mathbf t}$ is the isospin operator.

We have chosen to work with a parametrization which 
describes the properties of saturating nuclear matter proposed in 
\cite{Glen00}, since other common parameter sets, namely, TM1 \cite{tm1} and 
NL3 \cite{nl3} proved 
to be inadequate because, due to the inclusion of hyperons,  the nucleon mass 
becomes negative at relatively low densities. The chosen 
parameters are $g_s^2/m_s^2=11.79$ fm$^2$, $g_v^2/m_v^2=7.148$ fm$^2$,
$g_{\rho}^2/m_{\rho}^2=4.41$ fm$^2$, $\kappa/M=0.005896$ and 
$\lambda=-0.0006426$, for which the binding energy is -16.3 MeV at the 
saturation density $\rho_0=0.153$ fm$^{-1}$, the symmetry coefficient is 
32.5 MeV, the compression modulus is 300 MeV and the effective mass is $0.7 M$.
For the meson-hyperon coupling
constants we have opted for two sets discussed in the literature : set a) according  to \cite{gm91,Glen00} we choose the hyperon coupling constants 
constrained by the binding of the $\Lambda$ hyperon in nuclear matter, 
hypernuclear levels and neutron star masses
($x_\sigma=0.7$ and $x_\omega=x_\rho=0.783$) and assume that the couplings to the $\Sigma$ and
$\Xi$ are equal to those of the $\Lambda$ hyperon; set b) we take 
$x_{s B}=x_{v B}= x_{\rho B}=\sqrt{2/3}$ as in \cite{moszk,ghosh}. 
This choice is based on quark counting arguments.

The lagrangian density for the MIT bag model is identical to the one for the 
leptons, except for the degeneracy factor, which also accounts for the number 
of quark colors. In the energy density a factor $+B$ and in the pressure a 
factor $-B$ are inserted. This factor is responsible for the simulation of 
confinement. For the Bag model, we have taken B$^{1/4}$=190 MeV.

For the NJL model, the lagrangian density is
\begin{eqnarray}
L\,&=& \bar q\,(\,i\, {\gamma}^{\mu}\,\partial_\mu\,-\,m)\, q +\,g_S\,\,\sum_{a=0}^8\, [\,{(\,\bar q\,\lambda^a\, 
q\,)}^2\,\,+\,\,{(\,\bar q \,i\,\gamma_5\,\lambda^a\, 
q\,)}^2\,]\nonumber\\
&+&\  \,g_D\,\,  \{{\mbox{det}\,[\bar q_i\,(1+\gamma_5)\,q_j]
+ \mbox{det}\,[\bar 
q_i\,(1-\gamma_5)\,q _ j]\, }\},\label{1}
\end{eqnarray}
where $q=(u,d,s)$ are the quark fields and  
$\lambda_a$ $(\,0\,\leq\,a\,\leq\,8\,)$ are the
 U(3) flavor 
matrices. The model  parameters are: 
 $m\,=\, \mbox{diag}\,(m_u\,,m_d\,,m_s\,)$, the  current 
quark mass matrix ($m_d=m_u$), the 
coupling constants $g_S$ and $g_D$ and the cutoff in 
three-momentum space, $\Lambda$.

We consider the set of parameters \cite{Ruivo99,kun89}: 
 $\Lambda=631.4$ MeV, $ g_S\,\Lambda^2=1.824$,  $g_D\,\Lambda^5=-9.4$, $m_u=m_d=5.6$ MeV and
 $m_s=135.6$ MeV  which where fitted to the following properties: $m_\pi=139$
 MeV, $f_\pi=93.0$ MeV, $m_K=495.7$ MeV, $f_K=98.9$ MeV, $\langle\bar u u\rangle=\langle\bar d
 d\rangle=-(246.7\mbox{ MeV})^3$ and $\langle\bar s s\rangle=-(266.9\mbox{ MeV})^3$.

The condition of chemical equilibrium is  imposed through the  two 
independent chemical potentials for neutrons $\mu_n$ and electrons  
$\mu_e$ and it implies that the chemical potential of baryon $B_i$ is 
$
\mu_{B_i}=Q_i^B\mu_n-Q^e_i \mu_e,
$
where $Q^e_i$ and $Q_i^B$ are,
respectively, the electric and baryonic charge of baryon or quark $i$.
Charge neutrality implies $\sum_{B_i} Q^e_i \rho_{B_i} + \sum_l q_l \rho_l=0$ 
where $q_l$ stands for the electric charges of leptons. In the mixed phase 
charge neutrality is imposed  globally, 
$
\chi\, \rho_c^{QP}+ (1-\chi) \rho_c^{HP}+\rho_c^l=0,
$
where $\rho_c^{iP}$ is the charge density of the phase $i$, $\chi$ is the 
volume fraction occupied by the quark phase and $\rho_c^l$ is the electric 
charge density of leptons. We consider a uniform background of leptons in the 
mixed phase since Coulomb interaction has not been taken into account. 
According to the Gibbs conditions for phase coexistence, the baryon chemical 
potentials, temperatures and pressures have to be identical in both phases, 
i.e.,
$\mu_{HP,n}=\mu_{QP,n}=\mu_n,$ $\mu_{HP,e}=\mu_{QP,e}=\mu_e,$  $T_{HP}=T_{QP},\quad P_{HP}(\mu_n,\mu_eT)=P_{QP}(\mu_n,\mu_e,T),$
reflecting the needs of chemical, thermal and mechanical equilibrium, 
respectively.

If neutrino trapping is imposed to the system, the beta equilibrium condition
is altered to
$
\mu_{B_i}=Q_i^B\mu_n-Q^e_i (\mu_e-\mu_{\nu_e}).
$
In this work we have not included trapped muon neutrinos. Because of
the imposition of trapping the total leptonic number is conserved, i.e.,
$Y_L=Y_e+Y_{\nu_e}=0.4$. As already mentioned, neutrino trapping is important 
during the cooling of 
the protoneutron star. Hence, at $S=0$ ($T=0$), it is not an expected
mechanism.  Even 
though, we have chosen to show some results with neutrino trapping at $S=0$
only for the sake of comparison with results in the literature and with 
results for higher entropies.

In the following section the results obtained are shown and discussed.

\section{Results and discussion}

In all figures shown, unless stated otherwise, set a) for the meson-hyperon coupling constants was used.

\begin{figure}
\begin{center}
\begin{tabular}{cc}
\includegraphics[width=8.cm]{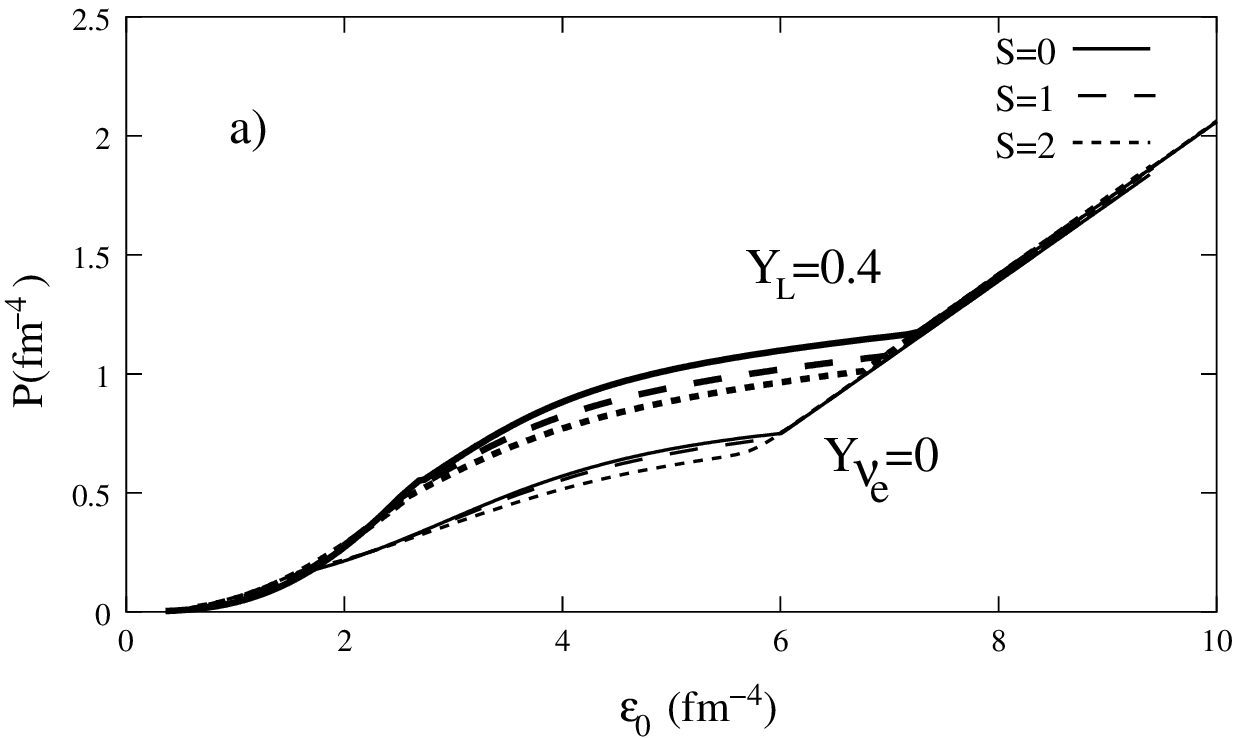}&
\includegraphics[width=8.0cm,angle=0]{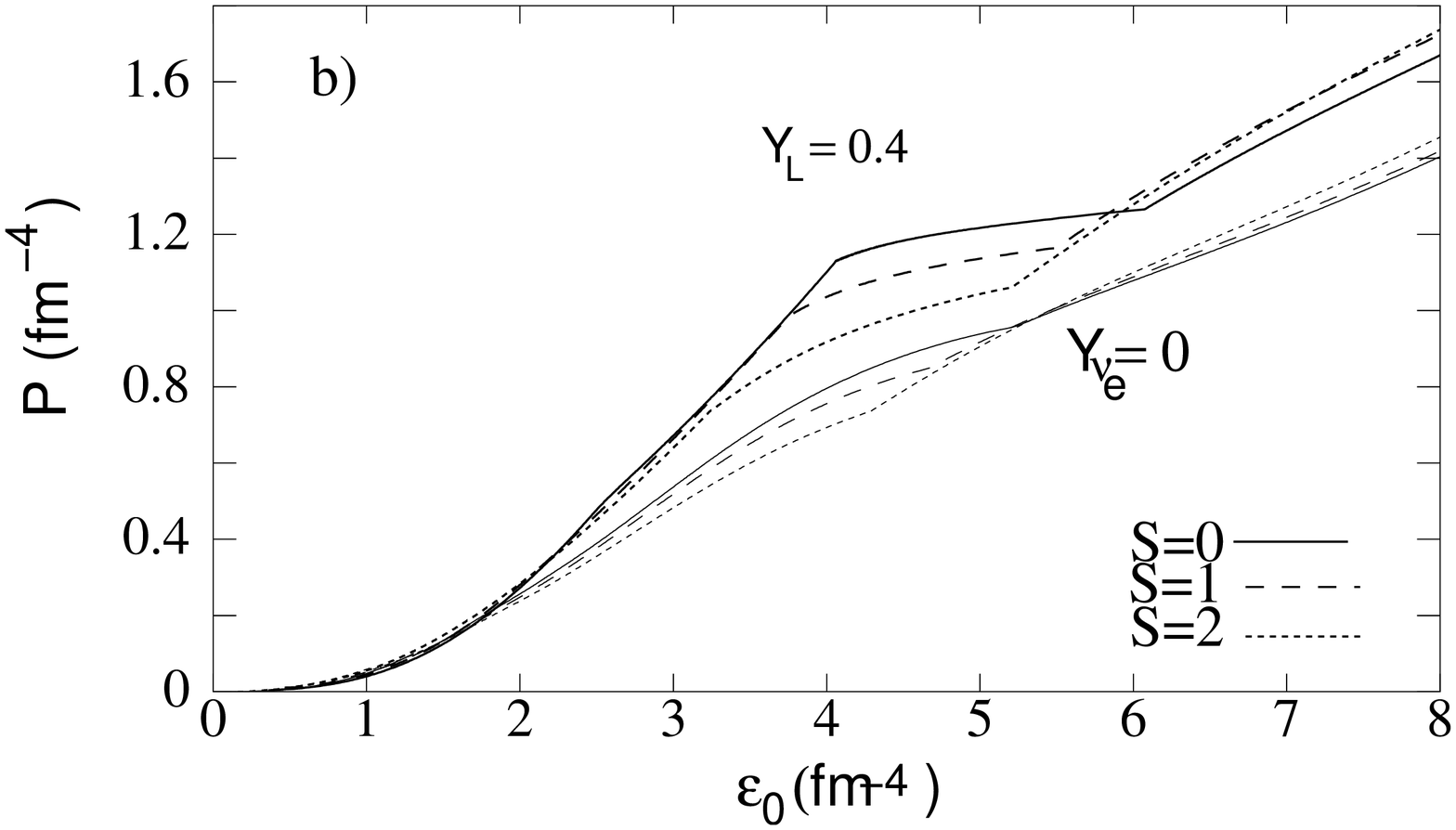}
\end{tabular}
\end{center}
\caption{EOS obtained with the GL force plus a) Bag model b) NJL model.}
\label{eos}
\end{figure}

In Fig. \ref{eos}, the EOSs obtained with both quark models are displayed for
$S=0$, $S=1$ and $S=2$ with neutrino trapping  ($Y_L=0.4$) and no neutrinos ($Y_{\nu_e}=0$). One can immediatly see that the EOS are harder 
and the mixed phase appears at higher densities if neutrino trapping is 
required independently of the model used. The energy density for the onset of 
the mixed phase can also be seen in Tables I to  IV. In general, the effect 
of temperature both in neutrino rich and neutrino poor matter is to decrease  
the density for the onset of the mixed phase. The only exception corresponds 
to neutrino poor matter obtained using the Bag model por the quark phase. 
This, however does not affect the maximum  mass of a stable star which in all 
cases decreases with increasing entropy, more strongly when the Bag model is 
used. An existing  difference between the EOS constructed with the NJL or the 
Bag model is the behavior of the respective EOS in the quark phase: contrary 
to the Bag model, the NJL model predicts an increase of the stiffness of the  
EOS both with increasing entropy and with neutrino trapping. However, this 
behaviour does not influence the properties of the compact stars because 
the calculated central densities of the most massive stellar configurations usually lie within the range of energy 
densities of  the mixed phase, as shown in tables I to  IV or are at most at
the borderline of the quark phase.

\begin{figure}
\begin{center}
\begin{tabular}{cc}
\includegraphics[width=8.0cm,angle=0]{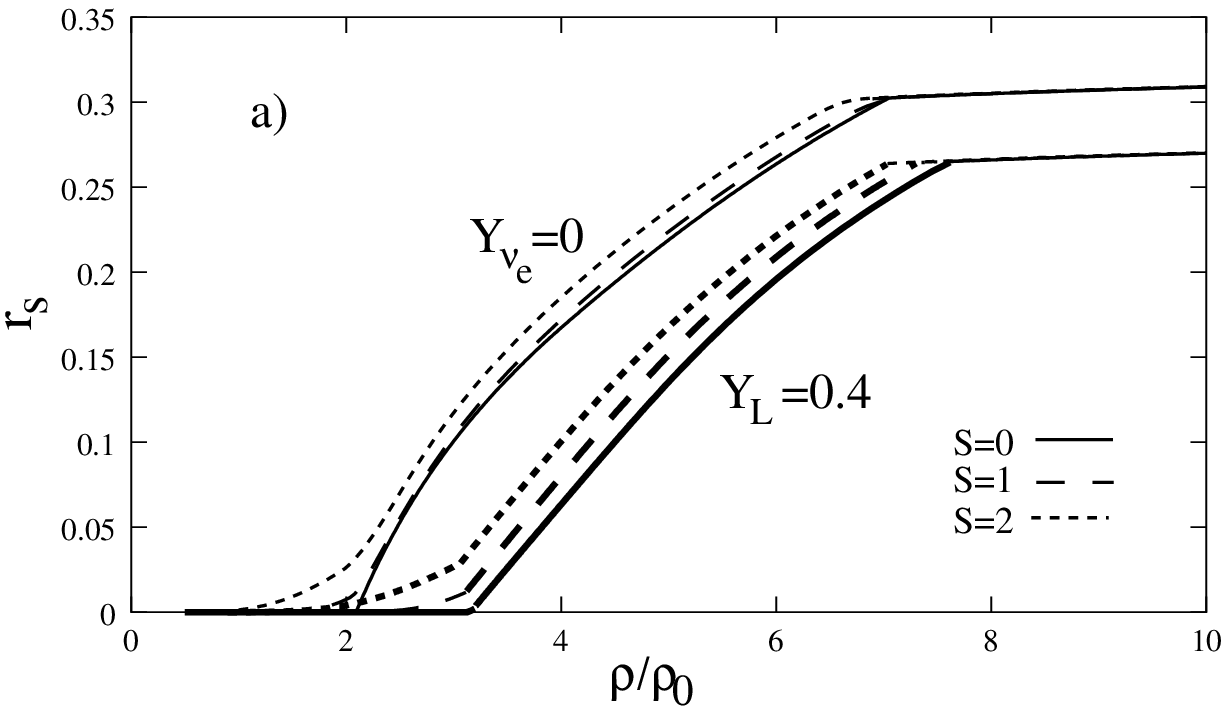}&
\includegraphics[width=8.cm,angle=0]{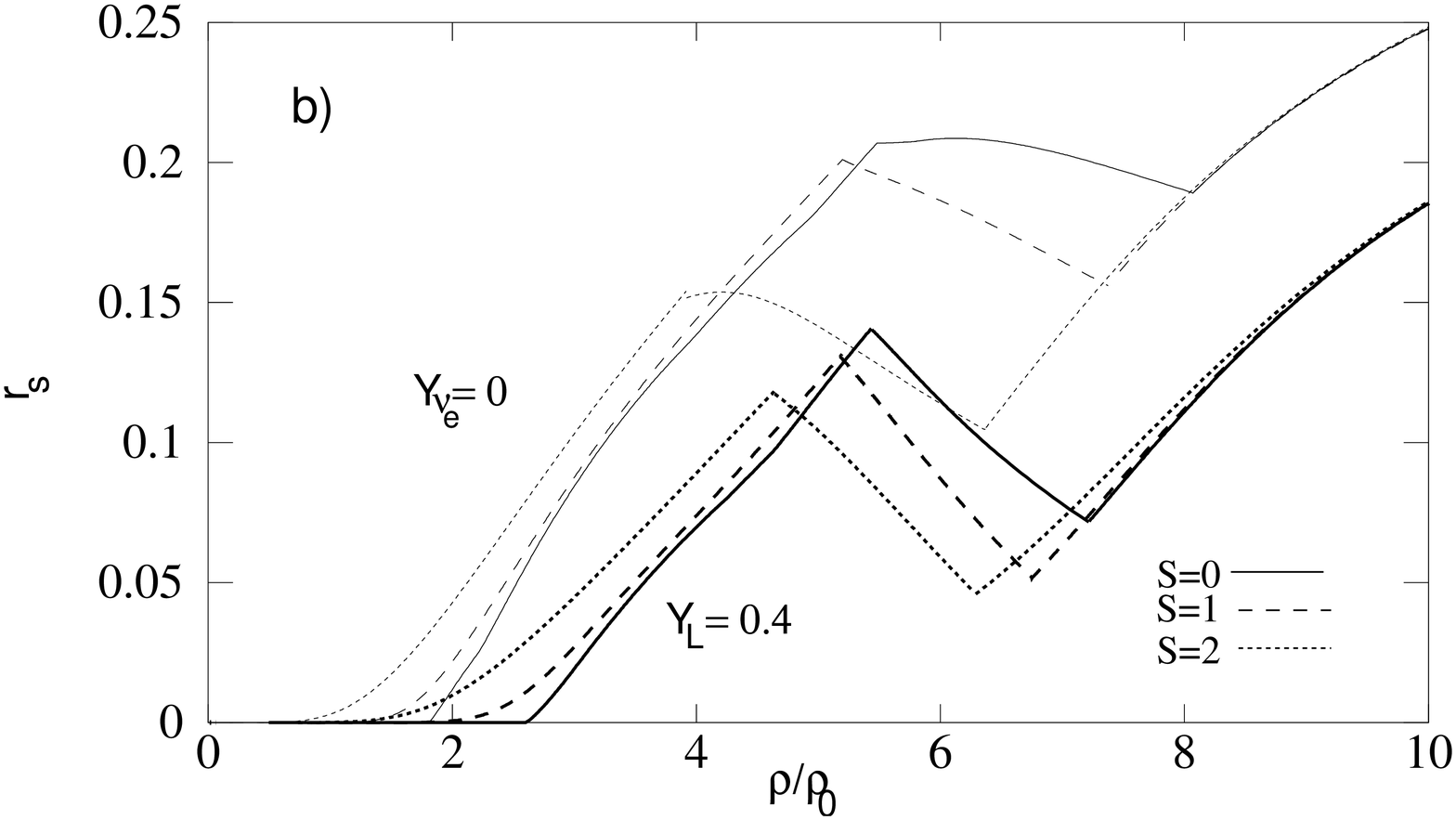}
\end{tabular}
\end{center}
\caption{Strangeness fraction $r_s$ for the EOS 
obtained with the GL force plus a) Bag model b) NJL model and set b).}
\label{schargenjl}
\end{figure}

As already discussed in \cite{recente}, the presence of strangeness in the 
core and crust of neutron and proto-neutron stars has important 
consequences in  understanding some of their properties. 
In Fig. \ref{schargenjl} we show, for the different 
EOSs, the 
strangeness fraction  defined as
$
r_s=\chi \, r_s^{QP}+ (1-\chi)\, r_s^{HP}
$
with $r^{QP}_s$ and $r^{HP}_s$ the quark and hadronic strangeness fraction, respectively.
For a  quark phase  described by the Bag model the strangeness fraction 
rises steadly and, at the onset of the pure quark phase it has almost reached  1/3 
of the baryonic matter if no trapping is imposed. However, it reaches a lower
value once neutrino trapping is enforced. This behavior is independent of the 
hyperon-meson coupling constants used in this work. The effect of the temperature
is to increase the strangeness fraction. 
The NJL model predicts a different behavior.  In the mixed phase the 
strangeness fraction decreases with density. This behavior is due to fact that for 
the densities at which the mixed phase occurs the mass of the strange quark is 
still very high.   In the hadron and mixed phase the strangeness fraction  depends on the parameter set employed and it is
larger for set b). 
In general,  temperature  increases 
the strangeness content and trapping decreases the strangeness
fraction.  
It is interesting to compare the Bag model and NJL model  results. At 10$\rho_0$ we get with the NJL model a
strangeness fraction of 0.25 for neutrino free matter and 0.18 with trapping. These numbers should be
compared with the results obtained with the Bag model, respectively 0.31 and 0.27.  This trend,
which is valid for all densities except for the mixed phase with 
set a), is due to the large $s$-quark mass in the NJL and has
consequences on the difference of the maximum baryonic
mass of the stars with and without neutrino trapping.

Analysing  Fig. \ref{schargenjl} we can also discuss the model
dependence of the hyperon and mixed phase onset.  
In general, there are larger fractions of  hyperons  if the NJL model
is used.  This is true for both the hadronic and the mixed phase and
is a consequence of the large s-quark mass in this model.
Trapping pushes the onset of hyperons, the mixed phase and the pure quark 
phase  to higher energies. The pure quark phase is only slightly affected but 
the mixed phase can occur
 at a density that is $1-2\rho_0$ higher.    For neutrino rich matter, the onset of
 hyperons  in the NJL model always occurs 
before the onset of the mixed phase  even for $T=0$, contrary to
neutrino free matter. The imposition of trapped neutrinos influences the threshold of 
hyperons and quarks through the conditions of charge conservation and chemical equilibrium. 
 
\begin{figure}
\begin{center}
\begin{tabular}{cc}
\includegraphics[width=8.0cm,angle=0]{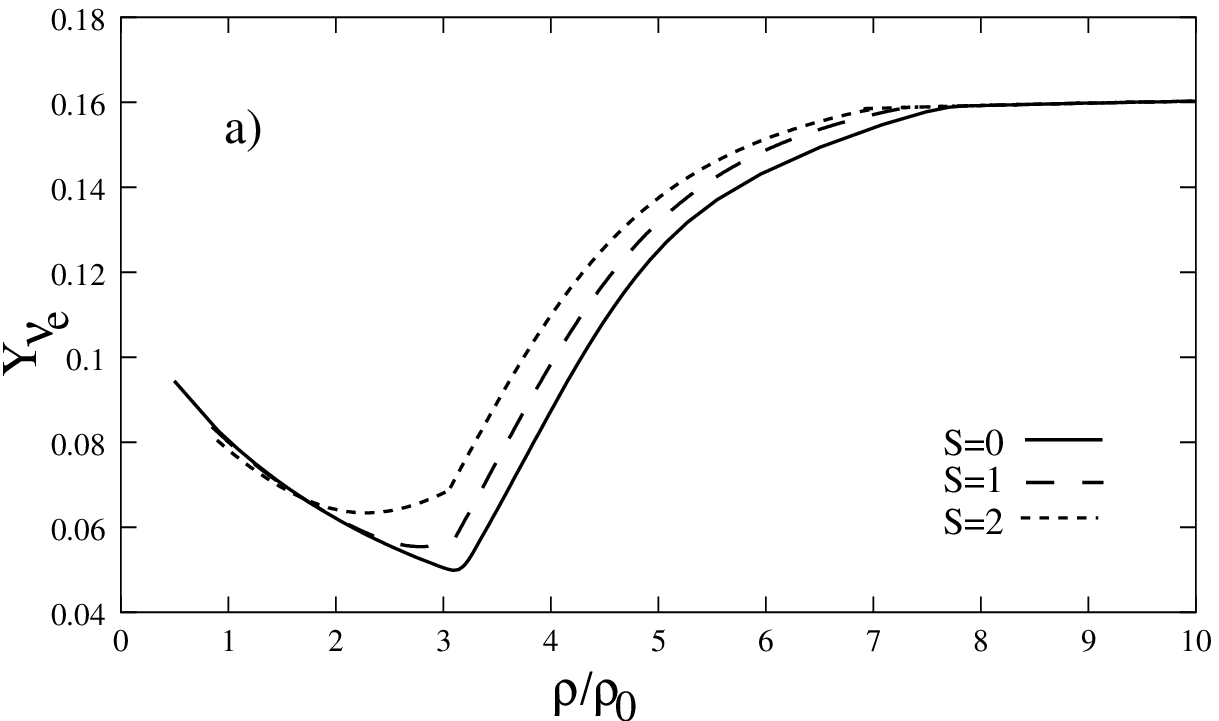}&
\includegraphics[width=8.0cm,angle=0]{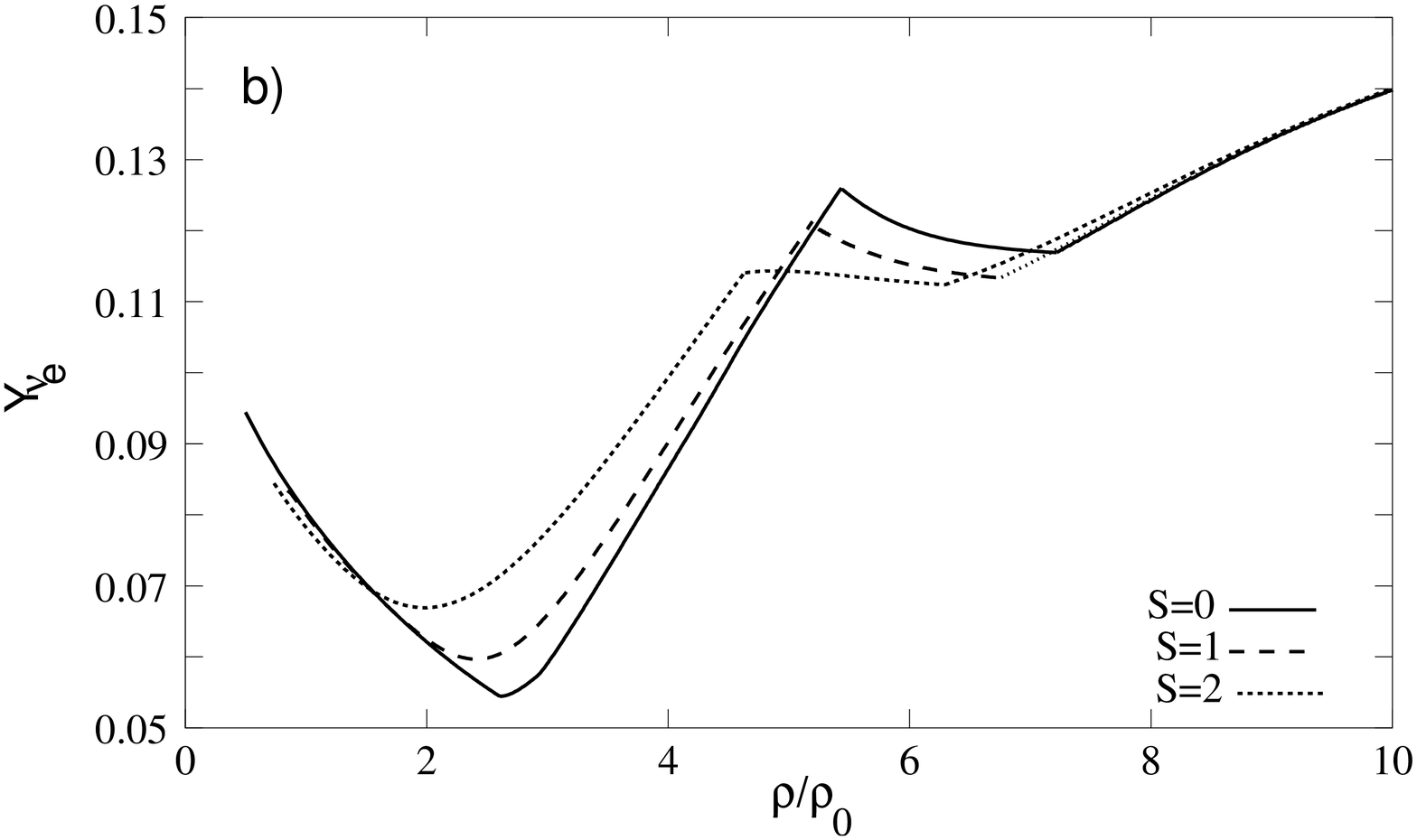}
\end{tabular}
\end{center}
\caption{Neutrino fraction for the EOS 
obtained with the GL force plus a) Bag model b) NJL model and set b).}
\label{neutrinonjl}
\end{figure}

In order to better understand the importance of the neutrinos when neutrino 
trapping
is imposed, in Fig. \ref{neutrinonjl} the fraction of
neutrinos is shown for increasing entropy. The behaviour encountered for the
Bag model (all parameter sets used) and the NJL model for set a) is quite 
similar. It decreases { at low densities when the electron fraction is
still increasing with density} and starts to increase
 in the hadron phase after the onset of hyperons, increases even more
 in the mixed and quark phases, where the fraction of
 electrons decreases continuously due to the onset of negative charged
 hyperons or the u and s-quarks.
In the mixed phase described within NJL for set b), the neutrino fraction decreases, due to
the decrease of hyperons and the very slow onset of the s-quark.  
In general, the
amount of neutrinos depends on the fraction of hyperons and quarks present in
each phase, which are determined by the model used. For any density 
the Bag model predicts higher neutrino yields. 

\begin{figure}
\begin{center}
\begin{tabular}{cc}
\includegraphics[width=8.cm]{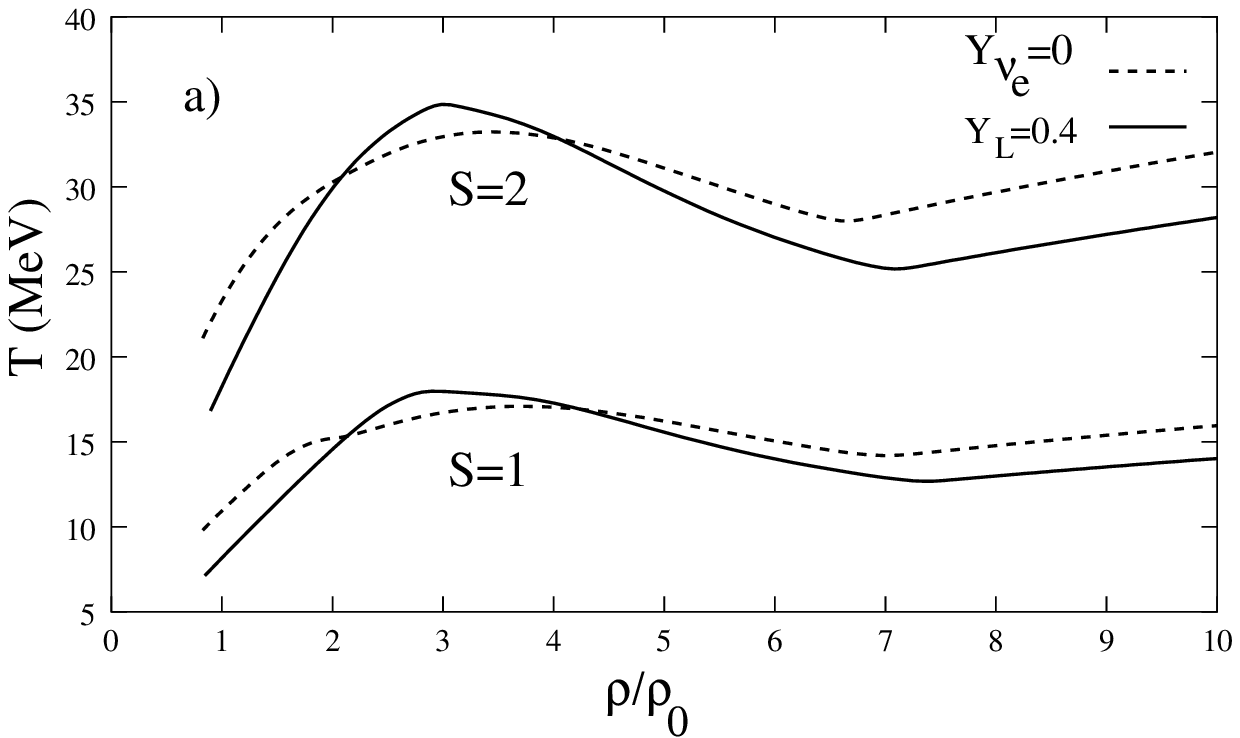}&
\includegraphics[width=8.50cm]{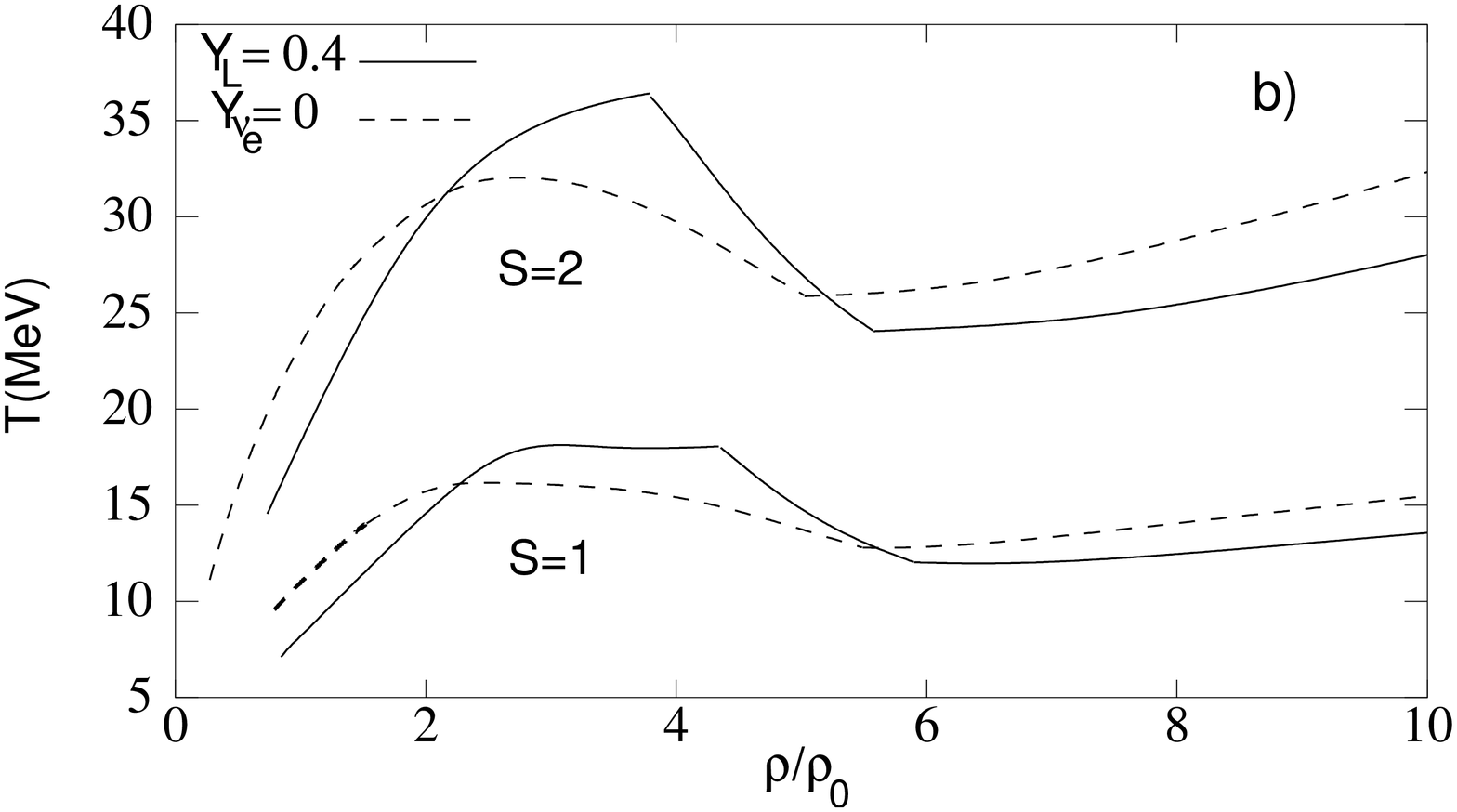}
\end{tabular}
\end{center}
\caption{Temperature range obtained with the GL force plus a) Bag model 
b) NJL model. In both figures the solid lines stand for the case with neutrino
trapping and the dashed line without neutrino trapping.}
\label{tpt}
\end{figure}

Finally, in Fig. \ref{tpt} the temperature is shown as a function of the 
baryon density for increasing entropy and for $Y_{\nu_e}=0$ and $Y_L=0.4$. 
The opening of new degrees of
 freedom has an important effect on the variation of temperature with density \cite{pons99}.  Neutrino
trapping makes the temperature vary more with baryonic density: temperature  attains a   higher value before
the onset of the  mixed
phase because trapping hinders the onset of hyperons, e. g. of new degrees of freedom; on the other
end temperature   becomes lower  in the quark phase because of the presence of  more degrees of freedom
, e. g. both leptons and quarks in opposition to the neutrino free matter which contains only
quarks. Both with or without  neutrino trapping  the mixed phase is
characterized by a decrease of the temperature. This behaviour,  a  colder high density EOS, is
due to deconfinement, and therefore to the appearance of a greater number of degrees of freedom. A  similar behaviour was obtained in \cite{spl00}. 

Mixed protoneutron and neutron star profiles can be obtained from all the EOS 
studied by solving the Tolman-Oppenheimer-Volkoff (TOV) equations \cite{tov}, 
resulting from a simplification of Einstein's general relativity 
equations for spherically symmetric and static stars. 
In Tables I, II, III and IV we show the values obtained for the maximum 
gravitational and baryonic masses of a 
neutron or protoneutron star as function of the central density for the EOSs 
studied in this work and for three fixed entropies. The results are shown for
the properties of the stars  with and without neutrino trapping. 
In Tables I and II the GL force and the NJL model were used to derive the 
full EOS, while in Tables III and IV the EOS was obtained from the GL force and the MIT Bag model.
Several conclusions can be drawn. For most EOSs studied, the central energy 
density $\varepsilon_0$ of the most massive stable stars falls inside the 
mixed phase, whose energy density
limits are shown as $\varepsilon_{min}$ and $\varepsilon_{max}$.
This is always true for the stars described with the MIT bag model. The 
results obtained within the NJL model depend on the hyperon couplings used: 
for set a) the core of the most massive stars is a pure quark phase, both for 
neutrino rich or neutrino free matter.
The maximum baryonic masses of the stars decrease with increasing entropy and are 
systematically larger if neutrino trapping is enforced. As a consequence, 
in the present
description the most massive stars with neutrino trapping are unstable after 
cooling. Similar results have already been discussed  for stars with strange
matter  \cite{prak97,vidana}.
Comparing Tables I, II and III, IV we conclude that the Bag model allows for smaller 
maximum gravitational  masses, of the order $\sim 1.9\,\, \mbox{ M}_{\bigodot}$ if neutrino
trapping is imposed and  $\sim 0.4\,\, \mbox{ M}_{\bigodot}$ lower otherwise, than 
the NJL model, $\sim 2.0 \,\, \mbox{ M}_{\bigodot}$ with neutrino trapping and
 $\sim 0.15\,\, \mbox{ M}_{\bigodot}$ lower  without it. This has been 
checked also for other Bag constants.
Comparing baryonic masses a similar conclusion is taken, i.e., within the Bag
model the most massive stable stars with neutrino trapping are
systematically  $\sim 0.4\,\, \mbox{ M}_{\bigodot}$ higher than the
corresponding neutrino free  stars. This difference reduces to  $\sim
0.1\,\, \mbox{ M}_{\bigodot}$ within the NJL model.

It is also seen that the maximum mass does not depend much on the hyperon 
couplings. One can also observe that the mixed phases start at higher 
densities if 
neutrino trapping is considered and they tend to shrink with increasing 
entropy. In general, changes in the maximum mass due to neutrino trapping are
larger than those due to variations in the entropy of the system.
Our results for the Bag model are systematically larger than those shown in 
\cite{prak97} where a bag constant equal to 197 MeV, corresponding to a harder
quark EOS, was used.
We do not discuss the radius of the maximum mass star because it is 
sensitive to the low density EOS and we did not describe properly this range 
of energy densities.

\section{Final Remarks}

In the present paper we have studied the EOS for proto-neutron stars using 
both the Bag model and
the NJL model for  describing the quark phase and a relativistic mean-field description in
which baryons interact via the exchange of $\sigma-,\, \omega-,\,\rho-$ mesons for the hadron phase. The EOS were constructed with and without the imposition of neutrino trapping at three fixed entropies.

For the hadron part of the EOSs we have considered a parametrization which 
describes the
properties of saturating nuclear matter proposed in \cite{Glen00}. For the 
hyperon meson coupling constants we
have used two choices and verified that for the NJL model the onset of the 
mixed phase and the properties of the corresponding  compact star are sensitive to the hyperon couplings.  

Some properties of the EOSs, as the strangeness fraction and the amount of 
neutrinos show different patterns depending on the quark model used, NJL or 
Bag model. Both strangeness and neutrino fractions are higher within the MIT bag model.
While the strangeness fraction  increases monotonicaly with  density for  the MIT bag
model, the NJL model predicts a decrease of the strangeness fraction
in the mixed phase, due to the large s-quark mass. 

The EOSs with trapped neutrinos are harder,  have smaller mixed phases which occur at higher
densities. These properties influence the properties of the corresponding compact stars: the
maximum baryonic allowed mass of a stable star is higher when
neutrinos are trapped. After cooling and deleptonization  these stars
will become unstable and decay into low mass blackholes
\cite{prak97,vidana}.  However,  it should be pointed out that the mass
difference is much smaller within the NJL model. In this model,   due to the only
partial chiral symmetry restoration the s-quark mass is still quite
high at the densities of interest for compact stars and therefore, the strange
content is smaller than the one obtained with the Bag model.

Another important characteristic of the EOSs studied  is the decrease of the 
temperature with density which occurs in the mixed phase. This is true for 
both models, NJL and bag model, and  for neutrino trapped or
 neutrino free matter. The temperature decrease is much stronger for matter 
with trapped neutrinos.

Within the present formalism, the core of the most massive stable stars, with few 
exceptions,  lies within the mixed phase, excluding the possibility of
stars with a quark core.
For the Bag model this fact is independent  of the hyperon coupling, contrary 
to what is observed with the NJL model: within this model the existence of 
stars with a quark core depends on the hyperon couplings chosen. 

For the quark phase we have chosen to use always unpaired quark matter.
Recently, many authors \cite{shh,ar,bo,arrw} have discussed the 
possibility that the quark matter is in a color-superconducting phase, in 
which quarks near the Fermi surface are paired, forming Cooper pairs which 
condense and break the color gauge symmetry \cite{mga}. At sufficiently high 
density the favored phase is called color flavor locked phase, in which quarks of all three colors and all three flavors are allowed to pair. The consequences
of using such paired quark phase in the construction of the EOS for the mixed 
phase are being investigated.

\section*{ACKNOWLEDGMENTS}
This work was partially supported by CNPq (Brazil), CAPES(Brazil)/GRICES(Portugal) under project 100/03 and FEDER/FCT (Portugal) under the project POCTI/35308/FIS/2000.

\newpage

\begin{center}
Table I - Hybrid star properties for the EOS obtained with the GL force and the
NJL model for fixed  entropies and with neutrino trapping ($Y_L=0.4$).
\end{center}
\begin{tabular}{lcccccc}
\hline
&S  & $M_{\mbox{max}}/M_{\bigodot}$&  $M_{\mbox{B~max}}/M_{\bigodot}$& 
$\varepsilon_0$ (fm$^{-4}$) & $\varepsilon_{min}$ (fm$^{-4}$) & $\varepsilon_{max}$ (fm$^{-4}$)\\
\hline

set a)        & 0 & 2.04 & 2.27 & 5.91 & 4.06 & 5.79 \\
$x_s=0.7$& 1 & 2.04 & 2.26 & 5.53 & 4.21 & 5.47 \\
$x_\omega=0.783=x_\rho$& 2 & 1.94 & 2.10& 5.26 & 3.25 & 5.20\\
\hline
set b) & 0 & 2.05 & 2.29 & 6.38 & 4.92 & 6.94\\
 $x_H=\sqrt{2/3}$                & 1 & 1.98& 2.19 & 6.08 & 4.72 & 6.46\\
                 & 2 & 1.96 & 2.12 & 5.68 & 4.50 & 6.00\\
\hline
\end{tabular}

\begin{center}
Table II - Hybrid star properties for the EOS obtained with the GL force and the
NJL model for fixed  entropies and no neutrino trapping  ($Y_{\nu_e}=0$).
\end{center}
\begin{tabular}{lcccccc}
\hline
&S   & $M_{\mbox{max}}/M_{\bigodot}$&  $M_{\mbox{B~max}}/M_{\bigodot}$& 
$\varepsilon_0$ (fm$^{-4}$) & $\varepsilon_{min}$ (fm$^{-4}$) & $\varepsilon_{max}$ (fm$^{-4}$)
\\
\hline
set a)  &0     &1.91& 2.20   & 4.90&1.30&5.21\\
$x_s=0.7$&1& 1.88  &2.13&4.80  &  1.09&4.66\\
$x_\omega=0.783=x_\rho$ &2& 1.82 &2.01 &4.66  &1.01&4.28\\
\hline
set b)&0 & 1.84& 2.09&6.26 &4.60&7.25\\
 $x_H=\sqrt{2/3}$&1  & 1.84 &2.09&  5.91 &4.35 &6.62\\
                 &2 &  1.82&2.02 & 5.33 &3.18 &5.66\\
\hline
\end{tabular}

\vspace{0.5cm}
\begin{center}
Table III -  Hybrid star properties for the EOS obtained with the GL force and 
the MIT Bag model for fixed entropies and with neutrino trapping ($Y_L=0.4$).
\end{center}

\begin{tabular}{lccccccc}
\hline
& S  & $M_{\mbox{max}}/M_{\bigodot}$ & $M_{\mbox{B~max}}/M_{\bigodot}$&
$\varepsilon_0$ (fm$^{-4}$) & $\varepsilon_{min}$ (fm$^{-4}$) & $\varepsilon_{max}$ (fm$^{-4}$) \\
\hline
set a)    & 0 & 2.00 & 2.22 & 5.06 & 2.73 & 7.38\\ 
$x_s=0.7$ & 1 & 1.91 & 2.07 & 4.95 & 2.52 & 6.95\\ 
$x_\omega=x_\rho=0.783$& 2 & 1.83 & 1.92 & 4.76 & 2.53 & 6.75\\ 
\hline
set b)            & 0 & 1.98 & 2.19 & 5.26 & 3.47 & 7.38\\ 
$x_H=\sqrt{2/3}$  & 1 & 1.93 & 2.00 & 5.06 & 3.03 & 6.99\\ 
                  & 2 & 1.81 & 2.19 & 5.35 & 2.98 & 6.82 \\
\hline
\end{tabular}
\vspace{0.5cm}
\begin{center}
Table IV -  Hybrid star properties for the EOS obtained with the GL force and 
the MIT Bag model for fixed entropies and no neutrino trapping ($Y_{\nu_e}=0$).
\end{center}
\begin{tabular}{lccccccc}
\hline
& S  & $M_{\mbox{max}}/M_{\bigodot}$ & $M_{\mbox{B~max}}/M_{\bigodot}$ &
$\varepsilon_0$ (fm$^{-4}$) & $\varepsilon_{min}$ (fm$^{-4}$) & $\varepsilon_{max}$ (fm$^{-4}$) \\
\hline
set a)    & 0 & 1.64 & 1.83 & 4.50 & 1.53 & 6.00\\ 
$x_s=0.7$ & 1 & 1.50 & 1.64 & 4.82 & 1.59 & 5.90\\ 
$x_\omega=x_\rho=0.783$& 2 & 1.50 & 1.62 & 4.60 & 1.67 & 5.73\\ 
\hline
set b)            & 0 & 1.64 & 1.83 & 4.50 & 1.81 & 6.06\\ 
$x_H=\sqrt{2/3}$  & 1 & 1.51 & 1.57 & 4.65 & 1.87 & 6.00\\ 
                  & 2 & 1.51 & 1.63 & 4.64 & 1.98 & 5.81 \\
\hline
\end{tabular}

\end{document}